\documentclass[%
superscriptaddress,
showpacs,preprintnumbers,
amsmath,amssymb,
aps,
prl,
twocolumn
]{revtex4-1}
\usepackage{ulem}
\usepackage{graphicx}
\usepackage{dcolumn}
\usepackage{bm}
\usepackage{color}

\begin {document}

\title{Machine learning prediction of self-assembly and analysis of molecular structure dependence on the critical packing parameter}

\author{Yuuki Ishiwatari}
\affiliation{%
Department of Mechanical Engineering, Keio University, Yokohama, Kanagawa 223-8522, Japan
}%

\author{Takahiro Yokoyama}
\affiliation{%
Department of Mechanical Engineering, Keio University, Yokohama, Kanagawa 223-8522, Japan
}%

\author{Tomoya Kojima}
\affiliation{%
Department of Applied Chemistry, Keio University, Yokohama, Kanagawa 223-8522, Japan
}%

\author{Taisuke Banno}
\affiliation{%
Department of Applied Chemistry, Keio University, Yokohama, Kanagawa 223-8522, Japan
}%

\author{Noriyoshi Arai}
\email{arai@mech.keio.ac.jp}
\affiliation{%
Department of Mechanical Engineering, Keio University, Yokohama, Kanagawa 223-8522, Japan
}%



\begin{abstract}
Amphiphilic molecules spontaneously form self-assembly structures based on physical conditions such as molecular structure, concentration, and temperature. These structures exhibit various useful functions according to their morphology. The concept of the critical packing parameter serves to correlate self-organized structures with chemical composition. However, unless both molecular arrangement and self-assembly patterns are understood, direct computational utilization for molecular design remains challenging.
In this study, we attempt to predict the self-assembled structure of a molecule directly from its chemical structure and analyze factors influencing it using machine learning. Dissipative particle dynamics simulations were used to reproduce many self-assembly structures composed of various chemical structures, and their critical packing parameters were calculated. A machine learning model was built using the chemical structures as input data and the critical packing parameters as output data.As a result, both Random Forest and a type of Recurrent Neural Network known as GRU demonstrated high predictive accuracy. It has been revealed through feature importance analysis and dependence on sample size that the amphiphilic nature of molecules significantly influences the self-assembly structures. Additionally, the importance of selecting an appropriate molecular structure representation for each algorithm has been emphasized.
The results of this research will help to further streamline product development in the fields of materials science, materials chemistry, and medical materials.
\end{abstract}

\maketitle

\section{Introduction}
Amphiphilic molecules have garnered significant attention as functional materials\cite{Domenico2015} and find applications in various fields such as materials chemistry\cite{Yan2023,Ambika2017,Shuo2019} and medical materials science\cite{Gunjan2013, Hu2018,Zhengyu2016}. Examples of such applications include detergents and liposomes used as drug delivery carriers\cite{Meel2014}, both of which exploit the functional properties of amphiphilic molecules. These functionalities are achieved through the self-assembly structures formed by the spontaneous aggregation of amphiphilic molecules. For example,  detergents utilise the properties of micelles whose hydrophobic tails adsorb to oils and trap oil stains within their hydrophobic cores. On the other hand, liposomes, used as drug carriers, encapsulate water-soluble substances in their interior and release them under specific conditions. In addition, other self-assembled structures, such as threadlike micelles\cite{Cao2023,Zonglin2013} and bilayer membranes\cite{Tien1998,Ming2022}, exhibit distinctive intrinsic properties that are being actively applied in various technological contexts. The design of such self-assembled structures for desired functions remains challenging as it depends not only on the chemical structure of the molecules but also on physical parameters such as concentration and temperature. As a result, there is no well-established recipe for achieving the desired functionality, leading to trial-and-error experiments that are time consuming and costly. Predicting and controlling the self-assembly of amphiphilic molecules is therefore a critical engineering task.

The concept that links the chemical structure of molecules to self-assembled structures is known as the critical packing parameter (CPP)\cite{Israelachvili2011}. This dimensionless quantity represents the geometric balance of hydrophilic and hydrophobic moieties at the interface of self-assembled structures. It is defined by the surface area of the hydrophilic part ($a_0$), the volume of the hydrophobic part ($v$), and the critical chain length ($l_c$) as $\text{CPP} = v / a_0 l_c$. The critical packing parameter plays a crucial role in determining the type of self-assembled structure formed. For example, when $0<\text{CPP}<1/3$, micelles form; $1/3<\text{CPP}<1/2$ results in thread-like micelles; $1/2<\text{CPP}<1$ leads to vesicles or flexible bilayer membranes; and $\text{CPP} \sim 1$, planar bilayer membranes are formed.  However, accurately calculating the critical packing parameter solely from the molecular structure poses challenges due to the influence of other thermodynamic conditions on self-assembled structures. Experimental estimation of the critical packing parameter has been attempted, but the results often deviate from the estimated values, necessitating a reliance on experimental observations to infer the structure. Therefore, an accurate calculation of the critical packing parameter requires consideration of both molecular shape and self-assembly information, making direct application to molecular design difficult. Consequently, to the best of our knowledge, there have been no reports of accurately predicting self-assembled structures using critical packing parameters estimated solely from molecular structure information.

On the other hand, machine learning is a technology that enables computers to iteratively learn from given data and uncover patterns within it, allowing predictions of unknown data. Artificial intelligence (AI) technology has found widespread applications in various fields, and in materials science, research using AI as materials informatics has been rapidly growing in recent years. Particularly, it shows a high affinity with soft matter and molecular simulation\cite{Ferguson2018,Friederich2019,Kadupitiya2020,Terao2020,Kim2021}. Inokuchi et al.\cite{Inokuchi2018} successfully predicted the properties of surfactants using a combination of molecular simulations and machine learning. This demonstrates the applicability of machine learning even in complex systems containing self-assembled structures of amphiphilic molecules. Bhattacharya et al.\cite{Bhattacharya2022} identified monomer sequences for self-assembling copolymers that form specific morphologies. However, these studies were limited to linear molecular models, leaving uncertainty about the feasibility of applying machine learning to molecular models of amphiphilic molecules with complex structures such as branching or cyclic structures.

This study focuses on the critical packing parameter, with the aim of predicting the self-assembled structure formed in water through the integration of machine learning and molecular simulation.
As previously highlighted, there have been no cases where the critical packing parameter has been accurately predicted without post-formation information about the self-assembled structure. Moreover, we meticulously scrutinize the input data for machine learning to identify the predominant factors affecting self-assembly. This enables us to forecast the resultant self-assembled structures without the requirement for trial-and-error experiments, offering the potential for a substantial contribution to the molecular design of functional materials and the advancement of materials science.

\section{Method}

\subsection{Molecular simulation}
We employed a simulation technique known as dissipative particle dynamics (DPD)\cite{Hoogerbrugge1992,Espanol1995,Groot1997,Santo2021} to study the self-assembly behavior of amphiphilic molecules. DPD has been specifically developed to simulate the fluidic and thermodynamic behavior of various aqueous solutions. Notably, it offers a computational advantage over classical molecular dynamics due to its efficiency. This efficiency is attributed to the fact that DPD is a particle-based method which uses coarse-grained models in which atoms and molecules are lumped together as DPD beads.

The fundamental equation in the DPD method is Newton's equation of motion, where three types of forces:conservative, dissipative, and random are applied to all DPD beads.
Newton's equation of motion for the particle $i$ is given as follows:

\begin{equation}
	m_i \frac{d {\textbf{v}_i}}{dt} = {\textbf{f}_i} = \sum_{j \neq i} {\textbf{F}}_{ij}^\mathrm{C} + \sum_{j \neq i} {\textbf{F}}_{ij}^\mathrm{D} + \sum_{j \neq i} {\textbf{F}}_{ij}^\mathrm{R}\;\;
  \label{eq:eq_motion}
  \end{equation}

where $m$ is the particle mass, $\textbf{v}$ is the particle velocity, $\textbf{F}^{\rm{C}}$ is the conservative force, $\textbf{F}^{\rm{R}}$ is the pairwise random force, and $\textbf{F}^{\rm{D}}$ is the dissipative force. The conservative force $\textbf{F}^{\rm{C}}$ is given by the following equation.

 \begin{equation}
	{\textbf{F}}_{ij}^\mathrm{C} =
		\begin{cases}
			-a_{ij} \left( 1-\dfrac{ \left| \textbf{r}_{ij}\right|}{r_{\mathrm c}} \right) \textbf{n}_{ij}, & \left| \textbf{r}_{ij} \right| \leq r_{\mathrm c} \\
			                      \;\;\;\;\;\;\;\;\;\;\;\;\;\;\;0,	& \left| \textbf{r}_{ij} \right| > r_{\mathrm c}\;\;,
		\end{cases}
	\label{eq:FC}
\end{equation}

\noindent where $\textbf{r}_{ij} = \textbf{r}_{j} - \textbf{r}_{i}$, and $\textbf{n}_{ij} = \textbf{r}_{ij} / \left| \textbf{r}_{ij} \right|$. $a_{ij}$ is the parameter that determines the magnitude of the repulsive force between particles $i$ and $j$, and $r_{\mathrm c}$ is the cutoff distance to determine the effective range of force.
The random force (${\textbf{F}}_{ij}^\mathrm{R}$) and dissipative force (${\textbf{F}}_{ij}^\mathrm{D}$) are given by the following equations.

\begin{equation}
	\label{eq:FR}
	{\textbf{F}}_{ij}^\mathrm{R} =
		\begin{cases}
			\sigma \omega^{\mathrm R} \left( \left| \textbf{r}_{ij}\right| \right) \zeta_{ij} \Delta t^{-1/2} \textbf{n}_{ij}, & \left| \textbf{r}_{ij} \right| \leq r_{\mathrm c}\\
			                      \;\;\;\;\;\;\;\;\;\;\;\;\;\;\;0,	& \left| \textbf{r}_{ij} \right| > r_{\mathrm c}
		\end{cases}
\end{equation}

\begin{equation}
	\label{eq:FD}
	{\textbf{F}}_{ij}^\mathrm{D} =
		\begin{cases}
			- \gamma \omega^{\mathrm D} \left( \left| \textbf{r}_{ij} \right| \right) \left( \textbf{n}_{ij} \cdot \textbf{v}_{ij} \right) \textbf{n}_{ij}, & \left| \textbf{r}_{ij} \right| \leq r_{\mathrm c},  \\
			                      \;\;\;\;\;\;\;\;\;\;\;\;\;\;\;0,	& \left| \textbf{r}_{ij} \right| > r_{\mathrm c}
		\end{cases}
\end{equation}

\noindent where $\bm{v}_{ij} = \bm{v}_{j} - \bm{v}_{i}$, $\sigma$ is the noise parameter, $\gamma$ is the friction parameter, and $\zeta_{ij}$ a random number based on a Gaussian distribution.
Here, $\omega^{\mathrm R}$ and $\omega^{\mathrm D}$ are $r$--dependent weight functions given as follows.

 \begin{equation}
  \label{eq:w_func}
	\omega^{D} \left( r \right) = \left[ \omega^{R} \left( r \right) \right]^2 =
		\begin{cases}
			\left[1 - \dfrac{\left| \bm{r}_{ij} \right|}{r_c}\right]^2,	& r_{ij} \leq r_c \\
			                      \;\;\;\;0,	& r_{ij} > r_c \;\;
		\end{cases}
\end{equation}

The temperature is controlled by a couple of dissipative and random forces.
The values of $\sigma$ and $\gamma$ are related by the fluctuation-dissipation theorem:

 \begin{equation}
  \label{eq:fd_theory}
	\sigma ^2 = 2 \gamma k_\mathrm{B} T,
\end{equation}

\noindent where $k_{\mathrm B}$ is the Boltzmann's constant and $T$ is the temperature.
In DPD simulations, reduced units are generally used.

In this study, we adopted the spring force ${\textbf{F}}_{ij}^\mathrm{S}$ defined by

\begin{equation}
	{\textbf{F}}_{ij}^\mathrm{S} =
			-k_{s} \left( 1-\dfrac{ \left| \textbf{r}_{ij}\right|}{r_{\mathrm s}} \right) \textbf{n}_{ij}
	\label{eq:FS}
\end{equation}

\noindent where $r_{s}$ is the equilibrium bond distance representing the bond between linked DPD beads in the modeled molecules, and $k_{s}$ is the spring constant.

\subsection{Simulation models and conditions}
\begin{figure}[tb]
    \centering
    \includegraphics[width=8cm]{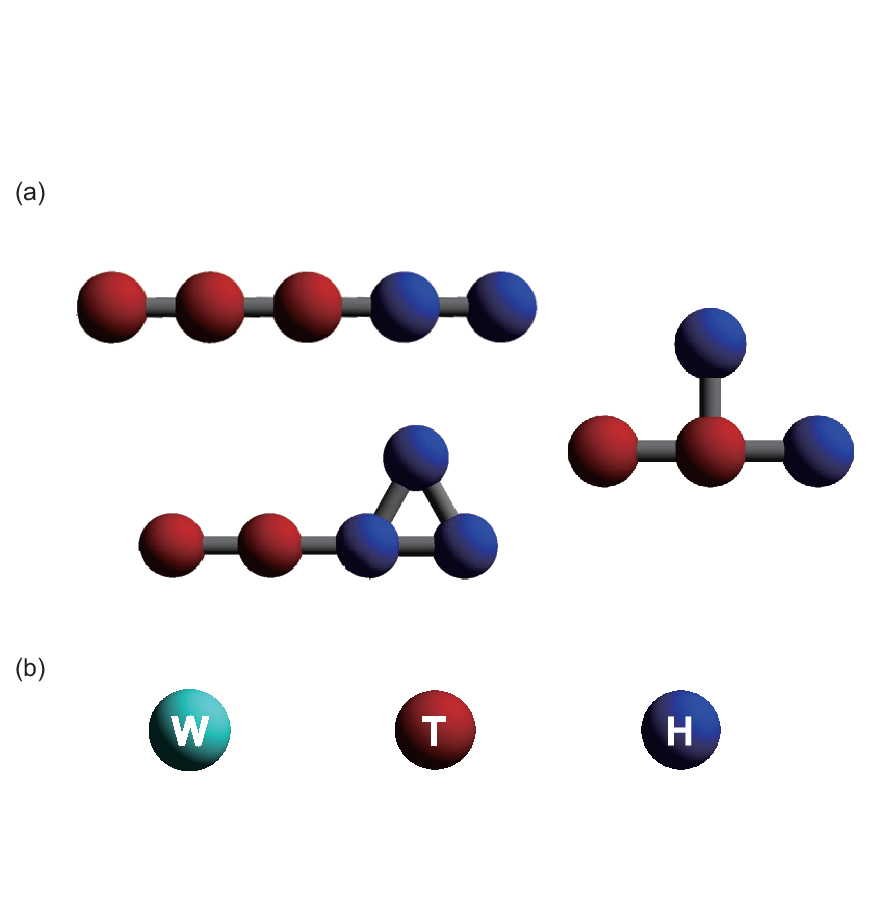}
    \caption{The particle model used for calculations. (a) Representative modeled amphiphilic molecules, which are composed of two types of DPD beads: hydrophobic tail (red) and hydrophilic head (blue). (b) Coarse-grained DPD beads used in this study. Water bead as solvent (cyan), hydrophobic tail bead (red), and hydrophilic head bead (blue) are labeled as W, T, H respectively.}
    \label{fig:model}
\end{figure}

A wide variety of amphiphilic molecule models were created by changing the number of coarse-grained particles and the arrangement of hydrophilic and hydrophobic groups.
Fig.\ref{fig:model} shows the simulation model used in this study. Typical amphiphilic molecular models are shown in Fig.\ref{fig:model} [a]. The amphiphilic molecule models include not only linear structures but also cyclic and branched structures. In total, more than 300 amphiphilic molecular models were used in the simulations. The water molecule (W) is represented as a single coarse-grained particle as shown in Fig.\ref{fig:model} [b]. The red particles in the amphiphilic molecular model represent hydrophobic groups (T), and the blue particles represent hydrophphilic groups (H). The interaction parameters ($a_{ij}$) between each pair of particles are shown in Table \ref{tbl:tbl1}. 

\begin{table}[tb]
\small
  \caption{\ Interaction parameters $a_{ij}$ (in $k_{\mathrm B}T/r_{\mathrm c}$ unit) in DPD calculations}
  \label{tbl:tbl1}
  \begin{tabular*}{0.48\textwidth}{@{\extracolsep{\fill}}llll}
    \hline
     & W & T & H \\
    \hline
    W & 25.0 & 75.0 & 25.0 \\
    T &      & 25.0 & 75.0 \\
    H &      &      & 25.0 \\
    \hline
  \end{tabular*}
\end{table}

The nearest neighboring particles within the modeled amphiphilic molecules are connected by harmonic springs with a spring constant of $100 k_{\mathrm B}T/r_{\mathrm c}^2$ and an equilibrium length of $0.86 r_c$. The concentration of the aqueous solution is set at 5\%. Specifically, the number of coarse-grained DPD beads in the modeled amphiphilic molecules and water molecules are 4050 and 76950, respectively. In this system, a random initial configuration was employed. The simulation box has a volume of $30 \times 30 \times 30 r_{\mathrm c}^3$, with periodic boundary conditions applied in all three dimensions. All simulations were conducted in a constant-volume and constant-temperature ensemble until reaching equilibrium state which took $16000\tau$.
In DPD simulations, typically a reduced unit system is employed. In this context, the length is given in terms of the cutoff distance $r_{\mathrm c}$, the mass employs the bead mass $m$, and energy is represented in units of $k_{\mathrm B}T$. The DPD time scale is defined as $\tau=r_{\mathrm c}(m/k_{\mathrm B}T)^{1/2}$. To correlate simulation results with real-world systems, a scaling procedure for length and time units is applied, following the approach introduced by Groot and Rabone\cite{Groot2001}. In this simulation, the coarse-graining of three water molecules into a single DPD particle results in a mass unit equivalent to 54 atomic mass units. The particle density in the simulation is set as $\rho r_{\mathrm c}^3 = 3$. This implies the inclusion of three DPD particles within the cube of $r_{\mathrm c}^3$, corresponding to a volume of $0.27 \text{nm}^3$, given that the volume of a water molecule is $0.03 \text{nm}^3$. Consequently, the physical value of the length unit $r_{\mathrm c}$ is $0.27^{1/3}$ $\text{nm}$ $ = 0.6463$ $\text{nm}$, and an approximate DPD time scale $\tau \approx 88 \text{ps}$ can be assumed.

\subsection{Machine learning}
In this research endeavor, we have utilized supervised learning in the form of regression analysis to predict self-assembled structures derived from a coarse-grained molecular model. To be more precise, we conducted a comprehensive assessment of a spectrum of well-established regression models. These encompass the likes of Lasso and Ridge, both of which represent variants of linear regression endowed with regularization techniques. Furthermore, we delved into regression models rooted in Support Vector Machine (SVM) frameworks, denoted as SVR (Support Vector Regression). Expanding our repertoire, we explored kernel-based methodologies such as the k-Nearest Neighbors(k-NN) approach. In addition, we explored ensemble techniques typified by the Random Forest algorithm. Lastly, we investigated broader regression frameworks exemplified by Neural Networks and a specific subclass known as Recurrent Neural Network (RNN).

In all instances, an ensemble of models underwent training via cross-validation using a dataset consisting of 305 samples. Within this framework, the structural data originating from coarse-grained molecular models were harnessed as input datasets, whereas the critical packing parameters served as the designated outputs. Nonetheless, a prerequisite for incorporating these coarse-grained molecular model structures as input lies in their transformation into a machine-readable format. In this investigation, we employed the Simplified Molecular Input Line Entry System (SMILES), a methodology adept at linearly encoding chemical structures, to facilitate the conversion of the structural data pertaining to the coarse-grained molecular models into an accessible form for our analyses.

Subsequently, the optimization of hyper parameters for each model becomes imperative. Given the limited extent of the dataset, we have opted for the methodical and encompassing grid search technique. The culminating parameter permutations have been ascertained to encompass those specific hyper parameters which yield the highest coefficient of determination (R2).

\subsection{Encoding for machine learning}
In the context of machine learning, handling chemical structures requires encoding them into a readable format. In this study, we performed the encoding of amphiphilic molecule models based on the SMILES notation.
The SMILES (Simplified Molecular Input Line Entry System)\cite{Weininger1988,Weininger1989,Weininger1990} is a widely-used method for converting chemical structures into linear notations, commonly employed when entering chemical information into databases. As implied by its name, it represents molecular information as a single string, making it easily understandable for both humans and computers. In SMILES, chemical structures are transformed into strings following specific rules. The key rules are outlined as follows:

\begin{enumerate}
    \setlength{\parskip}{0pt} 
    \item Atoms are denoted by their elemental symbols. Hydrogen atoms are usually omitted but can be explicitly indicated if necessary.
    \item Single bonds are implicit, and two adjacent atoms are automatically considered to be singly bonded. Double bonds are represented by "=", and triple bonds are represented by "\#".
    \item Branching structures are typically indicated using parentheses. For instance, the molecular formula of acetic acid is represented as CC(=O)O.
    \item Absolute configurations are denoted by "@" or "@@", while geometric isomerism is indicated using "/" or "\textbackslash".
    \item Ring structures are represented as broken chains, with the break points indicated by numbers. For example, cyclohexane can be written as C1CCCCC1.
    \item Aromatic rings are represented by lowercase letters for their constituting elements. As an illustration, benzene is written as c1ccccc1.
\end{enumerate}

\begin{figure}[htbp]
    \centering
    \includegraphics[width=8cm]{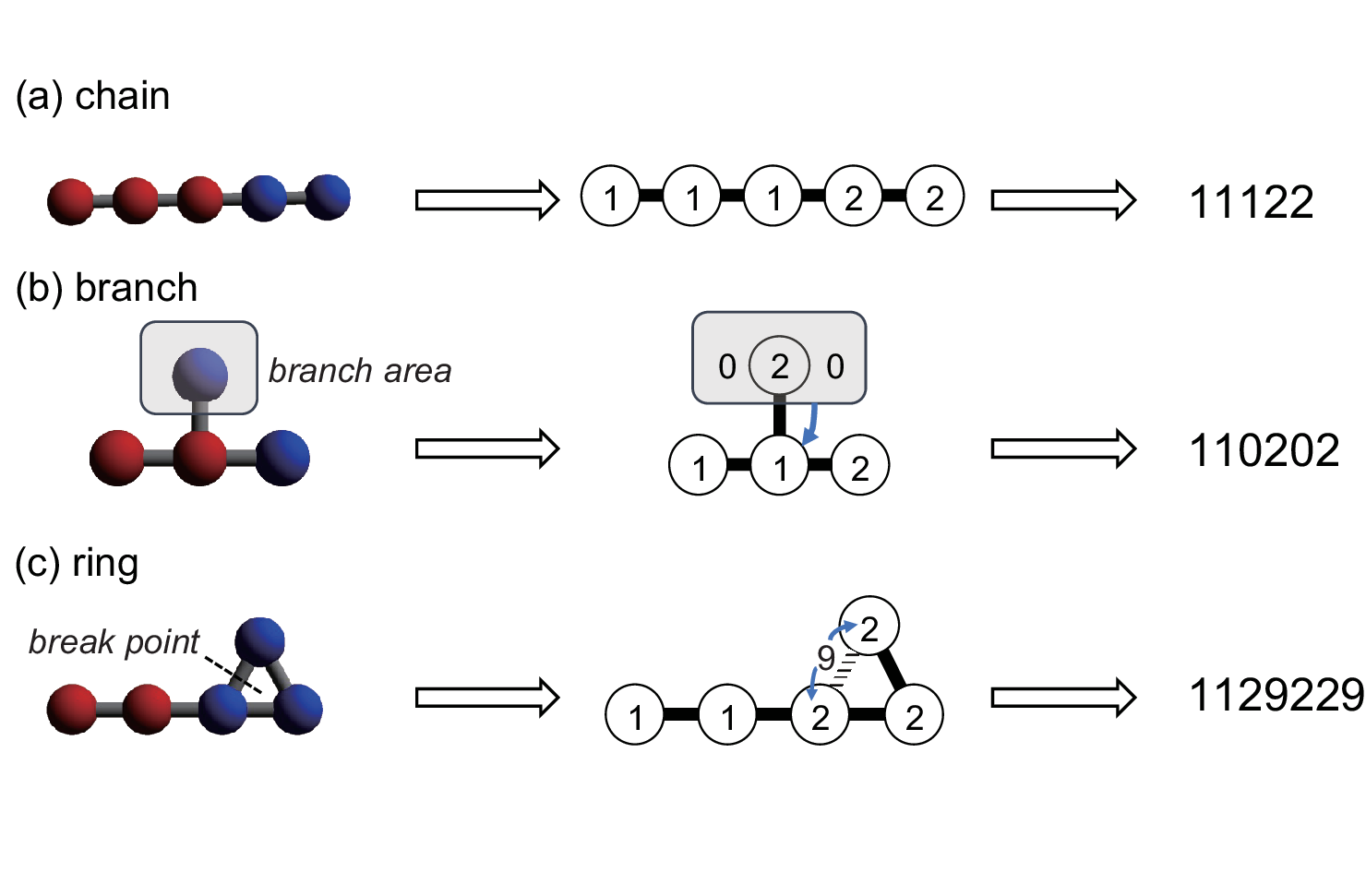}
    \caption{Modified-SMILES for linear transformation of coarse-grained molecular models.}
    \label{fig:smiles_example}
\end{figure}

\noindent The above rules are just a few examples, as there are more intricate details to accommodate the conversion of various structures.
In this research, we adopt a coarse-grained approach to model molecules, allowing us to represent the coarse-grained molecular model using simpler rules than conventional SMILES, while still capturing the necessary information adequately. We refer to this modified method as "modified-SMILES," and the rules for converting the coarse-grained molecular model into linear notation are as follows:
\begin{enumerate}
    \setlength{\parskip}{0pt} 
    \item Hydrophobic bead particles are represented by 1, and hydrophilic bead particles are represented by 2.
    \item Since all connections between particles have the same spring constant, all connections are represented by adjacent particle symbols.
    \item Branching structures are represented using 0 instead of parentheses.
    \item Similar to SMILES, ring structures are represented as broken chains, with the break points indicated by numbers. However, in this case, the break points are set to 9.
\end{enumerate}
\noindent By adhering to these rules, we can linearly represent the straight chains, branching, and ring structures in the coarse-grained molecular model. Specific examples are shown in Fig.\ref{fig:smiles_example}.

\section{Results and discussion}

\subsection{Analysis of self-assembly structures and the critical packing parameter}
\begin{figure}[tb]
    \centering
    \includegraphics[width=8cm]{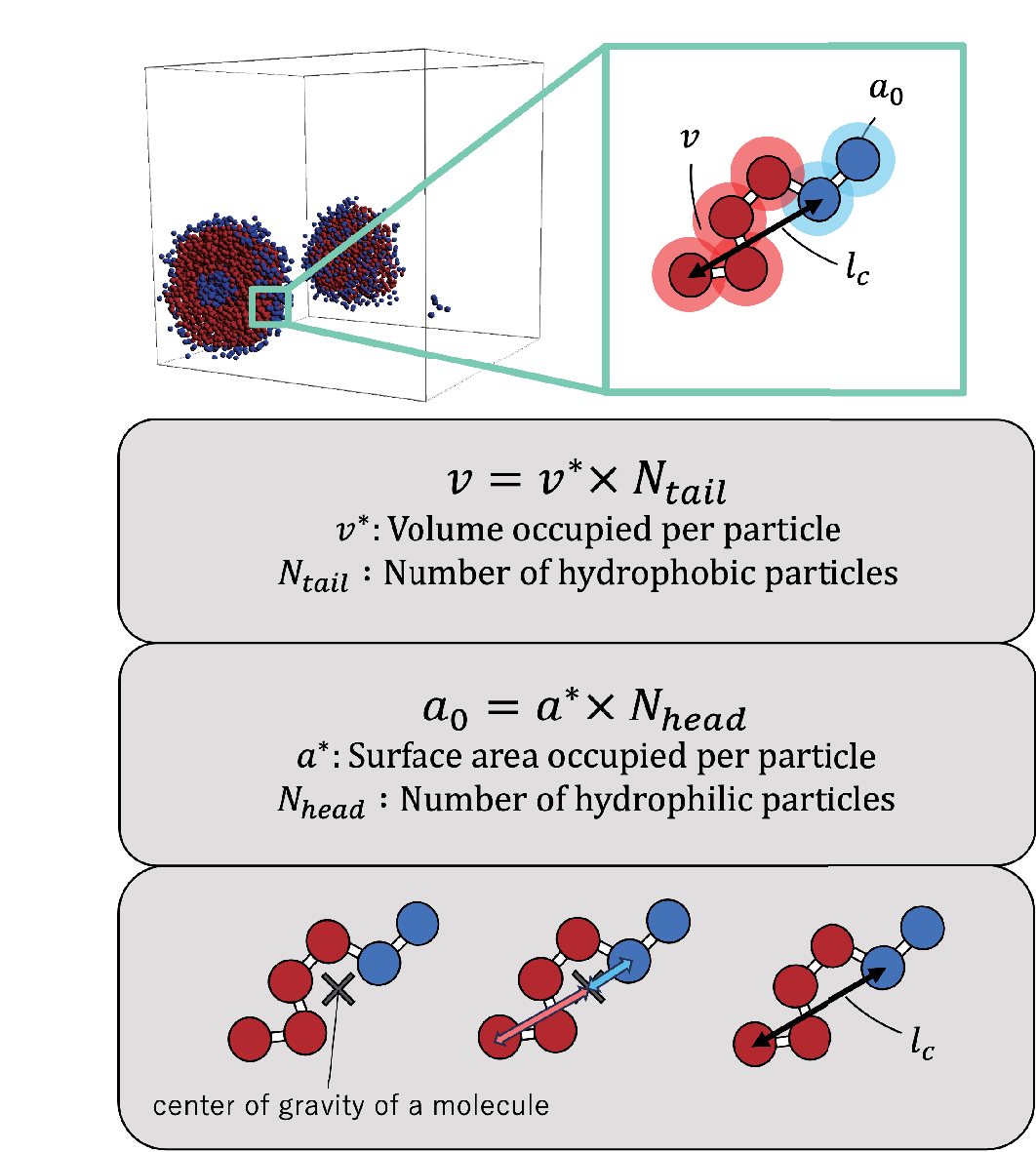}
    \caption{Calculation method for obtaining the three values needed to calculate critical filling parameters.}
    \label{fig:method_ana}
\end{figure}

\begin{figure}[tb]
    \centering
    \includegraphics[width=8cm]{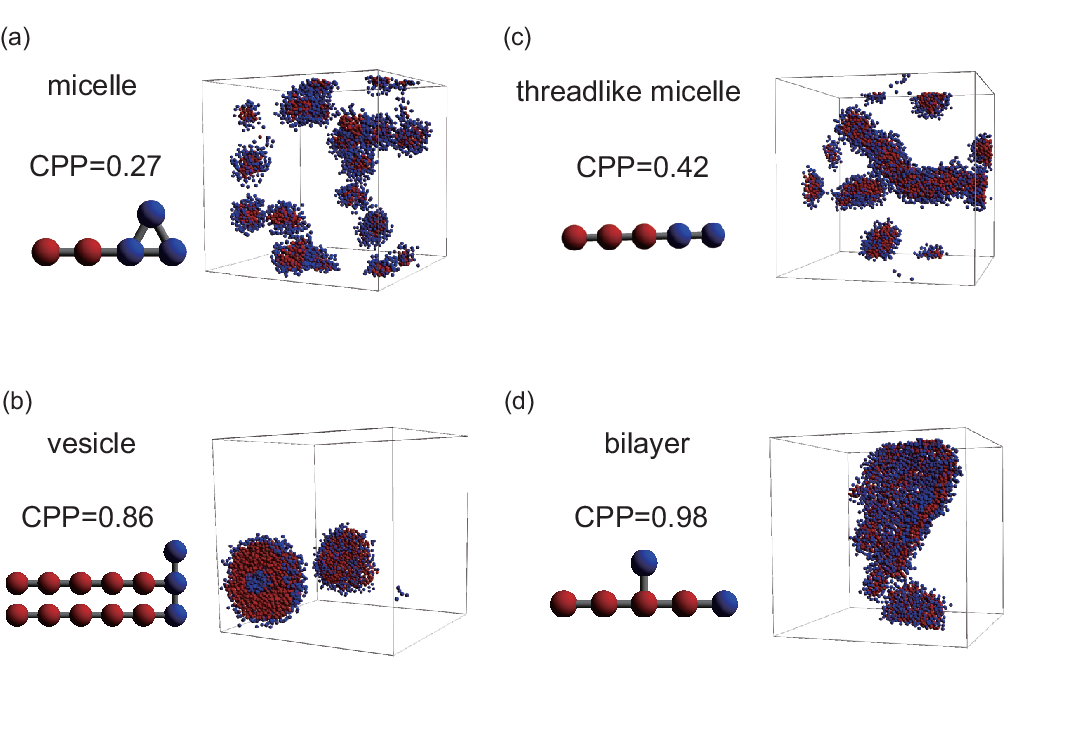}
    \caption{Representative snapshots of the equilibrium morphologies of amphiphilic molecules and the critical packing parameters of their models. (a) micelle, (b) vesicle, (c) threadlike micelle, (d) bilayer.}
    \label{fig:pp_selfassembly}
\end{figure}

As a result of the simulations, a variety of self-assembled structures, including micelles and vesicles, were observed, contingent upon the molecular structure. To determine the critical packing parameter ($\text{CPP}$) defined as below, we obtained values for the surface area of the hydrophilic portion ($a_0$), the volume of the hydrophobic portion ($v$), and the critical chain length ($l_c$) from the self-assembled structures generated during the simulations. 
\begin{equation}
    \label{eq:cpp}
    \text{CPP} = \frac{v}{a_0 l_c}
\end{equation}
\noindent Among the multiple clusters present within the system, we selected the self-assembled structure with the highest aggregation number as our target.
Fig.\ref{fig:method_ana} illustrates the method for calculating the three values necessary to determine the critical packing parameter from the simulation results. The volume ($v$) was computed by multiplying the volume occupied by a single hydrophobic particle by the number of hydrophobic particles within a single molecule. To obtain the volume per particle, we assumed a uniform particle distribution with a density of $\rho = 3$ and a constant number of particles within the system.
For the calculation of the surface area ($a_0$), we multiplied the number of hydrophilic particles in contact with water by the surface area per particle. The surface area per particle was derived by taking the two-thirds power of the volume per particle to perform the dimensional conversion.
The critical chain length ($l_c$) was defined as the distance between the center of mass of the molecule and the farthest hydrophobic particle, as well as the closest hydrophilic particle.
Using these computed values, we calculated the critical packing parameters for all the amphiphilic molecular models and verified the corresponding values for each self-assembled structure(Fig.\ref{fig:pp_selfassembly}).

\begin{figure}[tb]
    \centering
    \includegraphics[width=8cm]{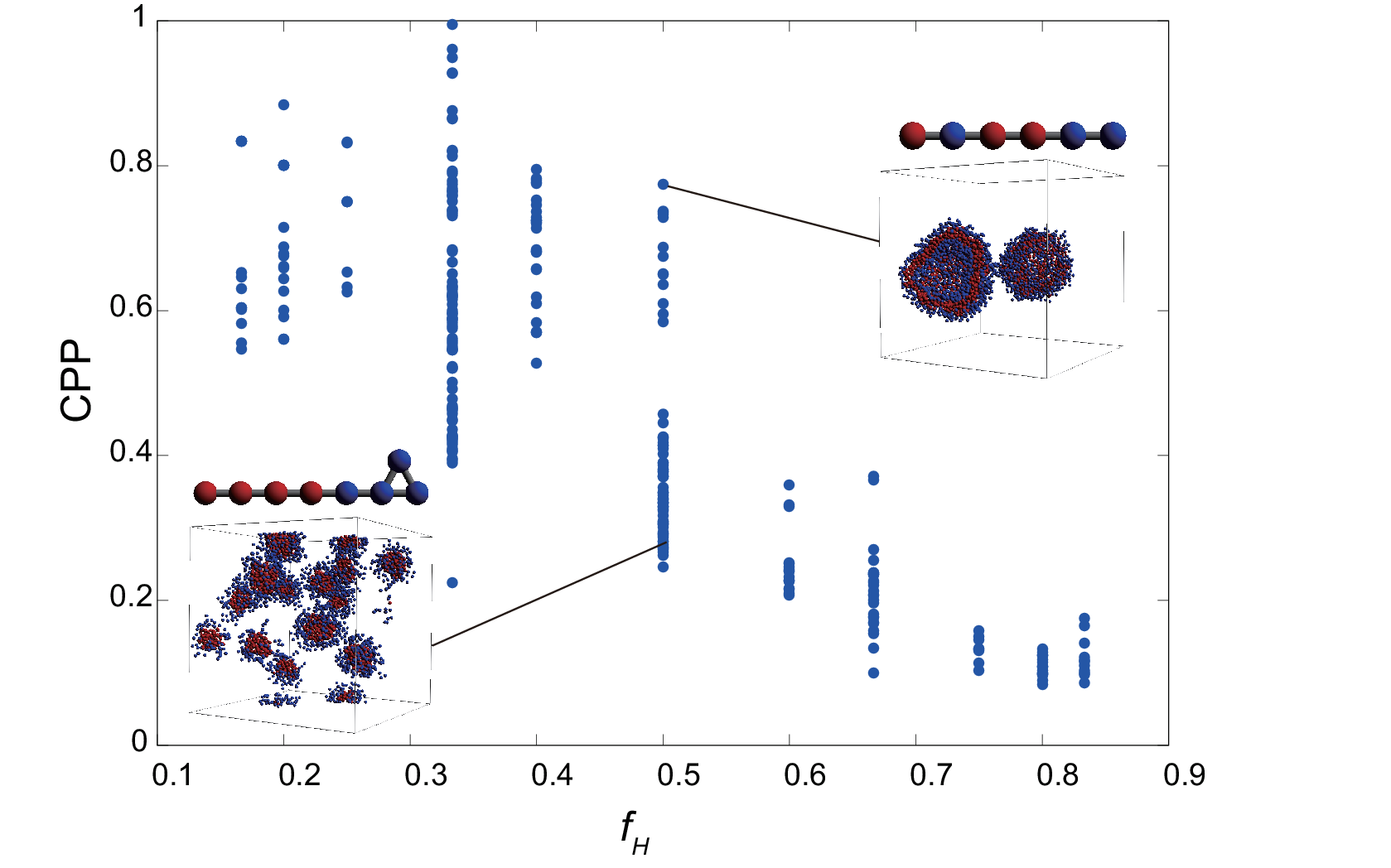}
    \caption{The scatter diagrams of the correlation between the critical packing parameter and the fraction of hydrophilic head groups.}
    \label{fig:head_percent}
\end{figure}

As the critical packing parameter of an aggregate derives from such as the surface area occupied by hydrophilic groups, and the volume encapsulated by hydrophobic moieties, it is to be expected that the relative abundance of hydrophilic and hydrophobic components within the molecule is anticipated to wield a pronounced influence on the resultant self-assembled configuration.
In fact, it has been studied that the proportion of hydrophilic or hydrophobic group of surfactant and diblock copolymer affects their self-assembly greately\cite{Mai2012,Lynd2008}.
In light of this context, we took a deliberate focus on the fraction of hydrophilic portion, and plotted a scatter diagram of critical packing parameter from simulations against the ratios of hydrophilic head particles within each molecule ($f_H$) in Fig.\ref{fig:head_percent}.
The figure reveals a notable negative correlation, with a coefficient of -0.826, between the critical packing parameter and the fraction of hydrophilic groups ($f_H$). This correlation can be attributed to the increased surface area ($a_0$) occupied by hydrophilic particles as their fraction rises.
However, even when the fraction of hydrophilic groups is kept constant, significant variations in self-assembled structures occur due to differences in factors such as the arrangement of particles, number of constituent particles, branching structures, and the presence of ring structures. This suggests that factors beyond the fraction of hydrophilic groups also play a crucial role in determining the self-assembled structures.
In fact, comparing at the same fraction of hydrophilic groups ($f_H=0.5$), differences in the location of ring structures and branching structures have been found to influence the value of CPP.
For example, if a molecule has a ring structure consist from hydrophobic tail beads, the critical chain length $l_c$ should be shorter than that of a molecule without cyclic structure, results in larger CPP. If a molecule has branches with hydrophilic head beads, the surface area of the hydrophilic portion $a_0$ will be bigger and that makes CPP smaller.
As such, the process of self-assembled structure determination should entail intricate interconnections among various conditions which makes it difficult to predict the self-assembled behavior from its molecular structure.
Thus, we utilized machine learning methods to analyze and discover key factors which rationalize the relationship between molecular structure and its self-assembly through the critical packing parameter.
This approaches will enable us not only to predict the self-assembly but to manipulate the self-assembly of amphiphiles through molecular design.

\subsection{Machine learning prediction}
We evaluated the performance of the following machine learning algorithm including deep learning methods: Lasso, Ridge, k-Nearest Neighbors (k-NN), Support Vector Regression (SVR), Random Forest, fully-connected Neural Network (NNs), Long-Short Term Memory (LSTM)\cite{Hochreiter1997}, and Gated Recurrent Unit (GRU)\cite{Cho2014}.
Lasso and Ridge regressions are regression techniques that add regularization to linear models to prevent overfitting and handle multicollinearity. The primary distinction stems from the utilization of distinct regularization methods. Lasso regression excels in the process of feature selection, effectively eliminating redundant attributes. In contrast, Ridge regression mitigates overfitting, thereby amplifying the overall stability of the model.
In this study, considering the limited dataset size of 305, linear regression models were also included for comparison. The other models are all non-linear.
k-NN is a regression algorithm that predicts the value of a new data point by considering the average (or weighted average) of the target variable of $k$ nearest neighbor data points. SVR is a regression model based on Support Vector Machines (SVM), which finds a hyperplane to predict the target variable while maximizing the margin with data points. SVR handles linear and non-linear relationships using kernel functions, making it suitable for high-dimensional data and complex regression tasks.
Random Forest is an ensemble learning method that introduces randomness to data and feature selection to construct multiple decision trees. In regression tasks, these tree predictions are combined to achieve accurate and robust predictions. It is known for its effectiveness in handling high-dimensional data, avoiding overfitting, and evaluating feature importance.
Neural Networks, inspired by the human brain, perform tasks like pattern recognition and prediction. Organized as layers of weighted nodes, they learn tasks through a learning process where weights are adjusted. Lastly, LSTM and GRU, both are types of Recurrent Neural Networks (RNNs), are useful models for processing time-series and sequence data. Generally, GRU is simpler and computationally efficient, making it suitable for small datasets and real-time applications. On the other hand, LSTM, being a more complex model, is well-suited for tasks with significant long-term dependencies.
Note that bi-directional RNNs algorithms were used in this study to learn the data of modified-SMILES more efficiently.

For evaluating the machine learning models' performance, we employed the coefficient of determination (R2) and the root mean square error (RMSE). The training dataset represents the data used for model learning, while the test dataset consists of unknown data and reflects the model's general performance. If there is a significant difference in performance between the training and test datasets, it indicates overfitting, where the model fits excessively well to the training data but lacks generalization capability.

\begin{figure}[tb]
    \centering
    \includegraphics[width=8cm]{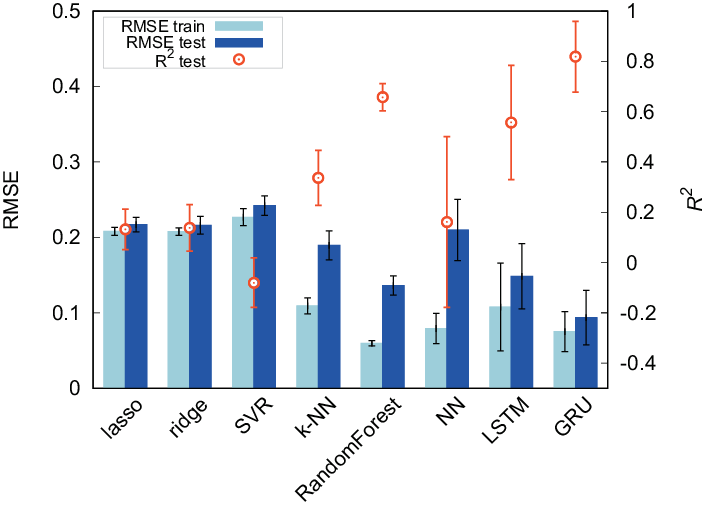}
    \caption{Comparison of model performance using modified-SMILES as explanatory variables, both root-mean-square error (RMSE) and model coefficient of determination ($R^2$) are shown. Here the error bars represent the standard deviation.}
    \label{fig:R2_RMSE_smiles}
\end{figure}

Firstly, we designated modified-SMILES as the explanatory variable and conducted an assessment of the model's performance, as depicted in Fig.\ref{fig:R2_RMSE_smiles}. Modified-SMILES encompasses comprehensive information pertaining to the hydrophilicity and hydrophobicity of the molecules, along with intricate details regarding branching structures and the presence of cyclic arrangements, thus serving as an all-encompassing representation of molecular structure. Consequently, utilizing modified-SMILES as the sole explanatory variable facilitates accurate predictions by the machine learning models.
The results demonstrate relatively high performance for random forest, LSTM, and GRU, as shown in Fig.\ref{fig:R2_RMSE_smiles}.
The heightened efficacy observed in LSTM and GRU models, in contrast to their machine learning counterparts, can be ascribed to the intrinsic characteristic of modified-SMILES as sequence data, which uses the order and presence of symbols to represent differences in molecular structures.
Except those RNNs algorithms, the order of data is not suggestive for predicting as in the previous study has already shown that the accuracy of random forest before and after the permutation of data didn't change much\cite{Bhattacharya2022}.
Hence, RNNs algorithms are suitable for those data such as molecular structure which is composed of sequential data.
However, compared to RNNs algorithms which need enormous number of data to create a decent model, Random Forest can be a good choice for the system with relatively small number of data such as this study with roughly 300 data.

\begin{figure}[tb]
    \centering
    \includegraphics[width=8cm]{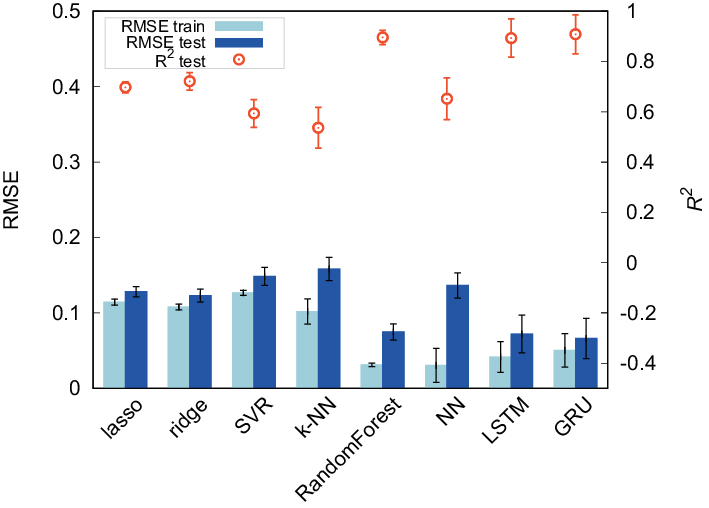}
    \caption{Comparison of model performance with explanatory variables set to modified-SMILES and supplementary values, both root-mean-square error (RMSE) and model coefficient of determination ($R^2$) are shown. Here the error bars represent the standard deviation.}
    \label{fig:R2_RMSE_all}
\end{figure}

While certain methods show high predictive accuracy, there remains scope for further improvement.
Therefore, in addition to the direct molecular structure representation, which is the modified-SMILES, we incorporate the frequency of $k$-mer tokens that represent the relative positional sense of different models.
Generally, a $k$-mer denotes a contiguous substring of length $k$ characters, and is often used to represent and analyze sequencial data such as DNA and amino acid sequences\cite{Rahman2018,Asgari2018}.
In this study, $k$-mers with $k$ under $2$ ($k=1,2$) are utilized since our models contain branch and ring structures so that we may not be able to capture the structures properly with the $k$-mers bigger than $k=3$.
In the context of this research, $k=1$ $k$-mers correspond to particle types (T, H), while $k=2$ $k$-mers represent combinations of adjacent particles (T-T, T-H, H-H), signifying distinct bond types.
Additionally to the counts of pure $k$-mer tokens, we came up with summing them up and calculating the fraction of them. The $k=1$ $k$-mers represent the number of each particle, so their sum becomes the overall number of particles in a molecule. The sum of $k=2$ $k$-mers represent the overall number of bonds which should reflect the molecular structure such as the number of branch and cyclic structures.
The ratio of particles or bonds against overall structures should also affect the self-assembly as shown in Fig.\ref{fig:head_percent}, so we calculated the fraction of $k$-mers by dividing the counts of each $k$-mers by their sums.
Thus, this study incorporates various explanatory variables, not only the counts of $k$-mers for $k=1$ and $k=2$ but also the sums of them and the fractions against the overall structures.
It is true that this $k$-mers derived data above is incorporated in modified-SMILES implicitly, however with the system with small number of data, such information may help to construct a better model to predict the self-assembly structures.
Hence, we used those $k$-mers derived data in conjunction with modified-SMILES representations to various machine learning algorithms to see the effects.
The assessment of model performance is illustrated in Fig.\ref{fig:R2_RMSE_all}. A comparison of R2 scores across models (random forest, LSTM, GRU) reveals that those models which leverage molecular structural insights derived from modified-SMILES exhibit comparatively heightened predictive accuracy. 
Moreover, across all machine learning models, augmenting modified-SMILES with $k$-mer information yields superior model performance compared to using modified-SMILES in isolation. This observation underscores the synergistic interplay between $k$-mer information and molecular structural insights from modified-SMILES.
In conclusion, the combined utilization of modified-SMILES and extracted particle and bond information emerges as an effective approach for predicting molecular structures and elucidating their properties. This study shows the importance of integrating diverse data sources to enhance the accuracy of predictive models.

\subsection{Machine learning analysis}
\begin{figure}[tb]
    \centering
    \includegraphics[width=8cm]{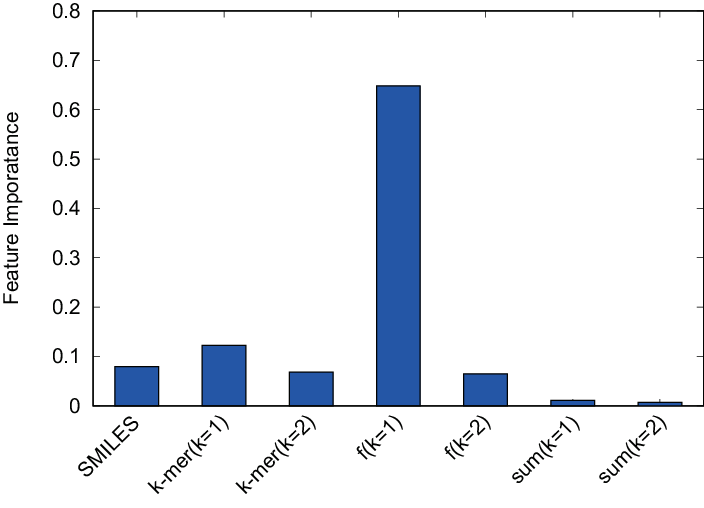}
    \caption{Feature importance of random forest model with modified-SMILES and supplementary values as explanatory variables.}
    \label{fig:feature_importance}
\end{figure}

Here, we contemplate the factors influencing the self-assembly structure determination. Firstly, an analysis is conducted regarding the explanatory variables inputted, based on the feature importance derived from the high-performance Random Forest in this study. Feature importance quantifies the contribution of each explanatory variable to predictions. The feature importance of each explanatory variable in the Random Forest predictions is illustrated in Fig.\ref{fig:feature_importance}. It is evident that the $k$-mer variables introduced play a significant role in predictions. While the particle count and bond count exhibited relatively lower values, $k$-mers, both $k=1$ and $k=2$, demonstrated feature importance comparable to that of SMILES. Moreover, the proportions of various $k$-mers ($k=1$), indicating the ratios of hydrophilic and hydrophobic particle components, contribute most prominently to predictions, underscoring their substantial involvement in self-assembly structure determination. Despite the indication from the heightened feature importance that the ratio of hydrophilic to hydrophobic particles dictates the self-assembly structure, Fig.\ref{fig:head_percent} has already demonstrated the inability of predicting self-assembly structures accurately based solely on the hydrophilic component ratio. This underscores the complementary roles played by each explanatory variable, signifying that determining the self-assembly structure of amphiphilic molecules necessitates the interplay of diverse factors. Among these, the proportion of hydrophilic and hydrophobic components within the molecule, essentially the amphiphilic nature, significantly influences the self-assembly structure of amphiphilic molecules.

\begin{figure}[tb]
    \centering
    \includegraphics[width=8cm]{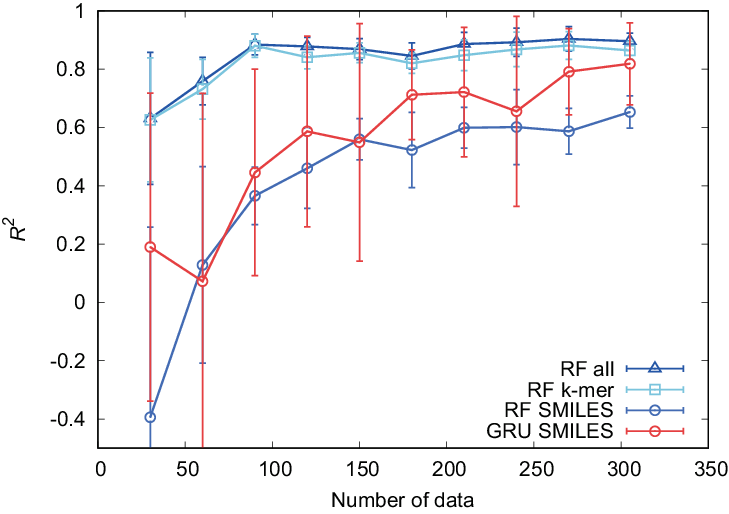}
    \caption{Sample size dependence of $R^2$ for three models: random forest and GRU with modified-SMILES as explanatory variables, and random forest with modified-SMILES and supplementary values as explanatory variables.}
    \label{fig:datasize}
\end{figure}

Next, in order to underscore the importance of the $k$-mer explanatory variables, we investigate the sample size dependence of prediction accuracy. The sample size dependence of prediction accuracy is illustrated in Fig.\ref{fig:datasize} for the cases where modified-SMILES and $k$-mer values are used as explanatory variables, as well as when only modified-SMILES or only $k$-mer are employed as an explanatory variable. 
Focusing on the random forest initially, it becomes evident that incorporating $k$-mer values as explanatory variables yields higher prediction accuracy across all sample sizes. Particularly noteworthy is the fact that when $k$-mer values are included as explanatory variables, high prediction accuracy is maintained even with smaller sample sizes. Reliable predictions can be achieved with sample sizes around 80. This not only accentuates the significance of the $k$-mer explanatory variables but also reinforces that the $k$-mer values introduced in this study are essential factors governing self-assembly structure determination.
This fact becomes evident when comparing the sample size dependency between utilizing modified-SMILES and $k$-mer as explanatory variables versus using only $k$-mer as the explanatory variable. In the case of Random Forest, incorporating $k$-mer as an explanatory variable can be deemed as contributing to the enhancement of predictive accuracy.

Moving on, we shift our attention to the sample size dependence of the GRU model employing only modified-SMILES as explanatory variables. At a sample size of 300, the model utilizing only modified-SMILES as explanatory variables demonstrates prediction accuracy equivalent to the random forest model, which incorporates both $k$-mer values and modified-SMILES as explanatory variables. This implies that the GRU can predict self-assembly structures from modified-SMILES alone, without the inclusion of $k$-mer explanatory variables. In other words, the GRU learns the fundamental factors determining self-assembly structures, such as the amphiphilic nature of molecules and molecular structural characteristics, directly from the modified-SMILES.
This fact suggests the potential for extending GRU-based self-assembly structure predictions to general molecular structures, not limited to coarse-grained molecules, by utilizing general SMILES. The absence of necessity to select and include auxiliary explanatory variables as additional inputs signifies the prospect of extending the predictions not only to self-assembly structures but also to various property predictions. However, it should be noted that when extending to general molecular structures, an increase in sample size is essential. Similar to the predictive results presented in Fig.\ref{fig:datasize} for the coarse-grained molecular model, improvements in predictive accuracy can be expected with increased sample sizes. Furthermore, extending to finer-grained general molecular structures requires a substantial increase in sample size.

To summarize the above, it is essential to appropriately select the representation method of molecular structures used as explanatory variables depending on the algorithm employed. GRU, by directly inputting the molecular structure representation, SMILES, is capable of learning the hydrophobicity of molecules and features of molecular structure, leading to highly accurate predictions. On the other hand, Random Forest cannot capture the fundamental factors determining self-assembly structures from SMILES, thus it is recommended to represent molecular structures using k-mer tokens and input them as explanatory variables.

\section{Conclusions}
We conducted a study utilizing molecular simulations and machine learning to investigate the relationship between the molecular structure of amphiphilic parent molecules and their self-assembly behavior. Various molecular structures including branch and cyclic structures were modeled, and their self-assembly structures were reproduced through molecular simulations.
To characterize the self-assembly structures, the critical packing parameter was adopted as a defining value. For all the reproduced self-assembly structures in the simulations, critical packing parameters were calculated. The critical packing parameters calculated in this study align with the actual self-assembly structures, serving as an evaluative index for self-assembly structures.
Furthermore, factors influencing self-assembly structures, such as the proportion of hydrophilic groups, were considered and indeed exhibited strong correlations. However, obtaining self-assembly structures (critical packing parameters) solely from molecular structures proved challenging based solely on these factors. Therefore, machine learning was employed to predict critical packing parameters and analyze factors influencing them.

The molecular structures were described using modified-SMILES as descriptors and input as explanatory variables. Initially, employing only modified-SMILES for machine learning, the results indicated that GRU achieved the highest predictive accuracy, followed by Random Forest and LSTM, which also demonstrated strong performance. Subsequently, incorporating k-mer into the machine learning process led to improved predictive accuracy across all algorithms.
Moreover, through feature importance analysis and sample size dependence, the study revealed that the amphiphilicity of molecules exerts the most significant influence on self-assembly structures. As different algorithms struggle to effectively learn this information, the selection of appropriate molecular structure representations for each algorithm becomes crucial.

Although this study maintained constant concentrations and temperatures, a future task involves creating models capable of accommodating variations in these parameters. Lastly, the results suggest that using RNNs allows for accurate prediction of self-assembly structures solely from SMILES, showcasing the potential for extending predictions of various material properties of amphiphilic molecules.
It was also shown that for the system with small number of data, exploiting random forest algorithm with scalar quantity such as $k$-mer tokens is beneficial for prediction.
Consequently, this holds the promise of significantly contributing to molecular design for functional materials in the field of materials science.
\section*{Conflicts of interest}
There are no conflicts to declare.

\bibliographystyle{ScienceAdvances}
\bibliography{./ref}

\end{document}